\documentclass[preprint,12pt]{elsarticle}

\usepackage{amssymb}
\usepackage{graphicx}
\usepackage{wrapfig}
\usepackage{url}
\usepackage{here}
\usepackage{tabularx}
\usepackage{amsmath}

\usepackage{multirow}

\journal{Nuclear Instruments and Methods in Physics Research, Section A (NIMA)}

\begin{document}

\begin{frontmatter}

\title{Fine residual stress distribution measurement of steel materials by SOI pixel detector with synchrotron X-rays}

\author[KEKIMSS]{Ryutaro NISHIMURA\corref{mycorrespondingauthor}}
\ead{ryunishi@post.kek.jp}

\author[KEKIMSS]{Shunji KISHIMOTO}

\author[KANAZAWAU]{Toshihiko SASAKI}

\author[KANAZAWAU]{Shingo MITSUI}

\author[KANAZAWAU2]{Masayoshi SHINYA}

\author[KEKIPNS]{Yasuo ARAI}

\author[KEKIPNS]{Toshinobu MIYOSHI}

\cortext[mycorrespondingauthor]{Corresponding author}

\address[KEKIMSS]{Institute of Materials Structure Science, High Energy Accelerator Research Organization (KEK-IMSS),\\ Oho 1-1, Tsukuba, Ibaraki, 305-0801, Japan}
\address[KANAZAWAU]{Institute of Human and Social Sciences, Kanazawa University, Kakumamachi, Kanazawa, Ishikawa 920-1192, Japan}
\address[KANAZAWAU2]{Graduate School of Natural Science and Technology, Kanazawa University, Kakumamachi, Kanazawa, Ishikawa 920-1192, Japan}
\address[KEKIPNS]{Institute of Particle and Nuclear Studies, High Energy Accelerator Research Organization (KEK-IPNS),\\ Oho 1-1, Tsukuba, Ibaraki, 305-0801, Japan}

\begin{abstract}

Residual stress is an important factor governing evaluating and controlling the quality of metal materials in industrial products. 
X-ray measurements provide one of the most effective means of evaluating residual stress without destruction. 
In such measurements, the effects of residual stress on the crystal structure can be observed through the Debye ring deformation. 

In previous studies, we developed a residual stress measurement system based on the $cos \alpha$ method, using a two-dimensional (2D) silicon-on-insulator pixel (SOIPIX) detector known as INTPIX4. 
In a typical laboratory setup, this system requires only 1 second to measure a specified point. 
This is drastically faster than the conventional system based on the $sin^{2} \psi$ method, which requires more than 10 min, 
and the $cos \alpha$-based system using an imaging plate, which requires 1 min. 
Compared to other systems, it can evaluate the 2D distribution of residual stress faster and provide more detailed information for evaluating materials. 
We first attempted to measure the 2D distribution in a laboratory setup with a Cr X-ray tube (Cr K$\alpha$ 5.4 keV) and obtained satisfactory results. 
We subsequently took measurements using synchrotron monochromatic X-rays to determine the fine accuracy and fine sampling pitch distribution. 
In this paper, we report the results of the initial synchrotron experiment, including the residual stress distribution of the standard specimen obtained by the first prototype setup. 
Furthermore, we compare the synchrotron measurements with those from the laboratory. 

\end{abstract}


\end{frontmatter}

%%\linenumbers

\section{Introduction}

\begin{figure}[htb]
\centering
\includegraphics[width=\linewidth]{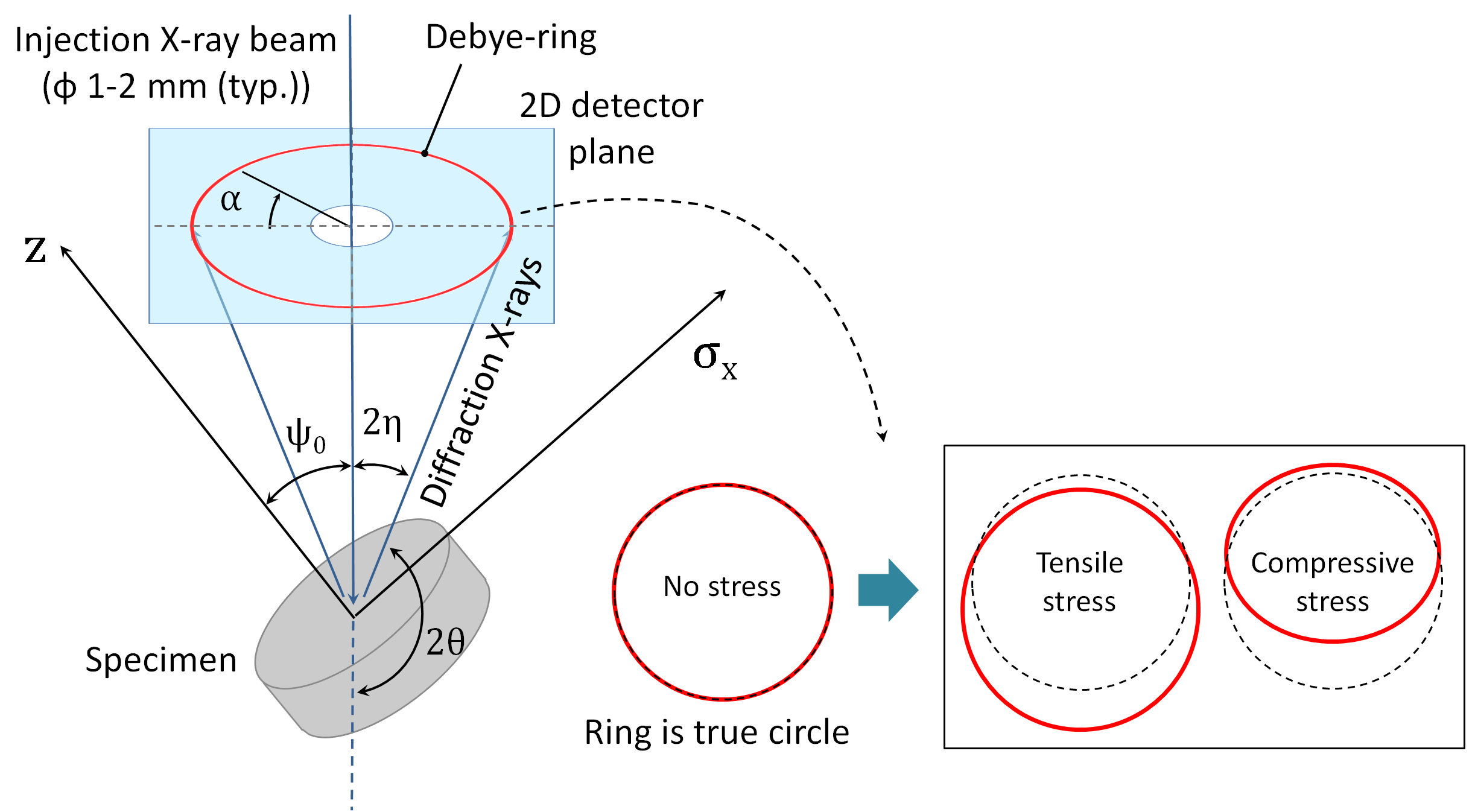}
\caption{Schematic of $cos \alpha$ method.}
\label{fig:cosa}
\end{figure}

Residual stress is an important factor governing the evaluation and control of the quality of metal materials in industrial products. 
Measurements using X-rays provide one of the most effective methods for evaluating residual stress without destruction. 
In such measurements, the diffraction of X-ray beams on the surface of a polycrystalline metal forms a Debye ring. 
The effects of residual stress on the crystal structure can be observed through the Debye ring deformation (Fig. \ref{fig:cosa}). 

A two-dimensional (2D) X-ray detector is required to measure the shape of the Debye ring. 
 We used a silicon-on-insulator (SOI) pixel detector known as INTPIX4 \cite{soi3, soi2} in our previous studies \cite{cosa, cosa2}. 
The SOI pixel detector is a new type of monolithic X-ray radiation image sensor that uses SOI technology \cite{soi}. 
It is a Si semiconductor detector with a small pixel size that can detect the charges generated by X-rays. 
Compared with hybrid X-ray detectors \cite{eiger, medipix3}, which use mechanical bump bonding to connect the large-scale integration (LSI) circuit and sensor, 
the SOI pixel detector can achieve higher spatial resolution (smaller than 10 $\mathrm{\mu m}$, depending on the scale of the circuit in each pixel). 
In our previous studies, we had developed a residual stress measurement system based on the $cos \alpha$ method \cite{cosa3, cosa4, cosa5} using INTPIX4. 
Moreover, we had attempted to measure the 2D distribution in a laboratory setup using a Cr X-ray tube (Cr K$\alpha$ 5.4 keV) and obtained satisfactory results. 
In this study, as the next step, we took measurements using synchrotron monochromatic X-rays to determine the fine accuracy and fine sampling pitch distribution. 

\section{Residual stress measurement system based on $cos \alpha$ method using INTPIX4}

\subsection{SOI detector}

\begin{figure}[htb]
\centering
\includegraphics[width=\linewidth]{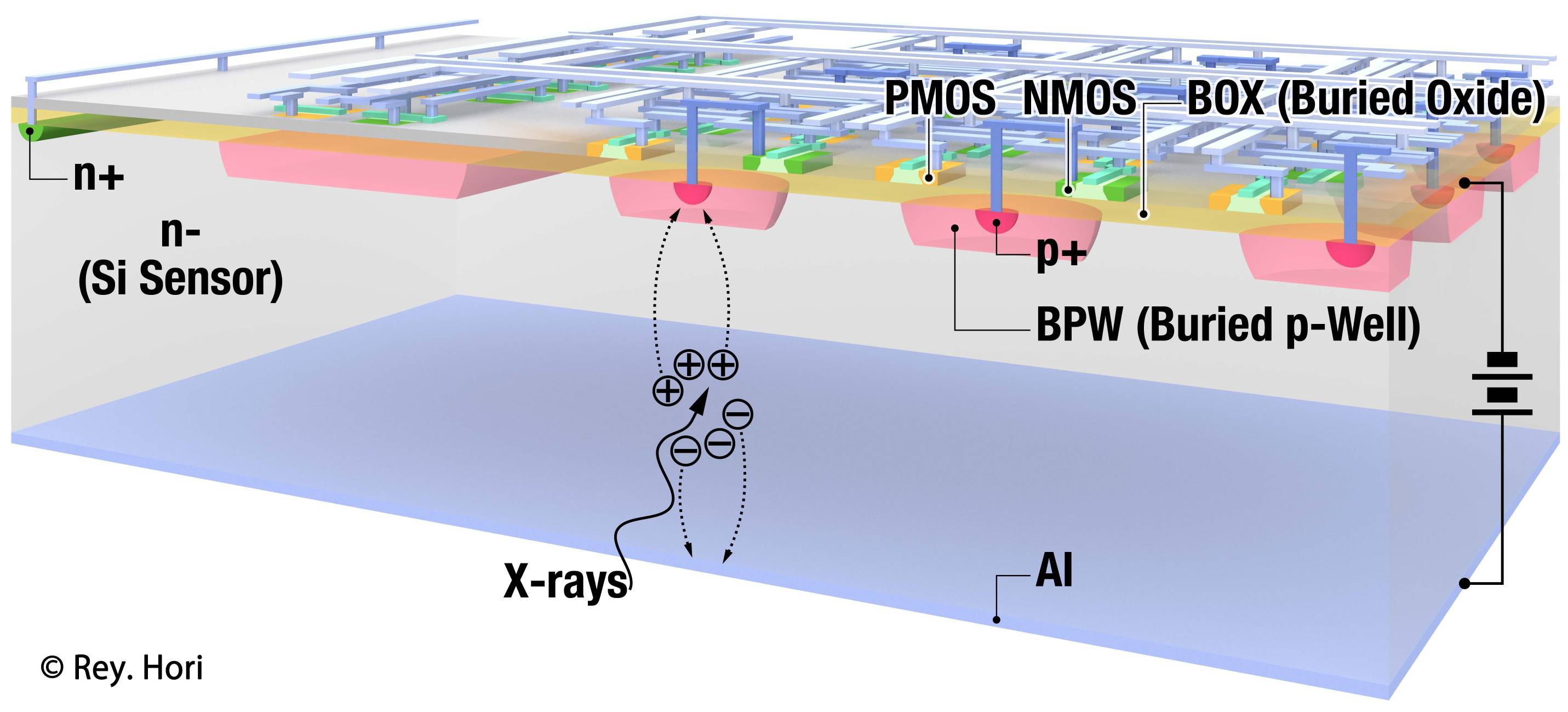}
\caption{Structure of SOI detector.}
\label{fig:soistruct}
\end{figure}

SOI pixel detectors are currently being developed by an SOI pixel (SOIPIX) collaboration led by KEK (the High Energy Accelerator Research Organization, Tsukuba, Ibaraki, Japan). 
These detectors are based on a 0.2 $\mathrm{\mu m}$ CMOS Fully Depleted Silicon On Insulator (FD-SOI) pixel process developed by Lapis Semiconductor Co., Ltd \cite{soi}. 
An SOI detector consists of a thick, high-resistivity Si substrate for the sensor and a thin Si layer for the CMOS circuits (Fig. \ref{fig:soistruct}). 
An SOI detector has no bump bonding; therefore, the application has low parasitic capacitance (\verb|~|10 fF), is effective for low noise and high conversion gain, and has a low material budget. 
Furthermore, it can operate rapidly with low power. 

\subsection{Overview of INTPIX4}

\begin{figure}[htb]
\centering
\includegraphics[width=\linewidth]{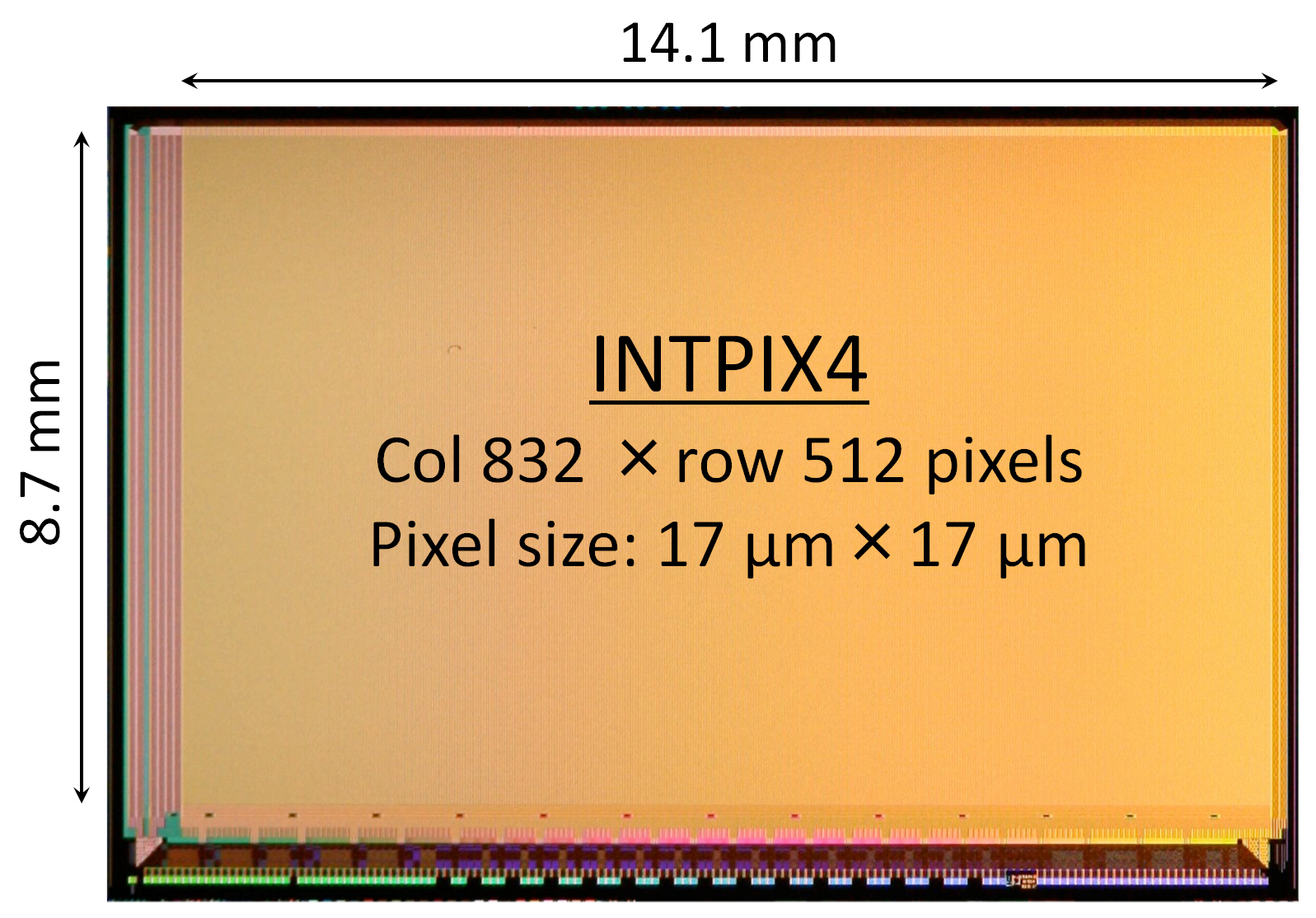}
\caption{Photograph of INTPIX4.}
\label{fig:intpix4}
\end{figure}

INTPIX4 \cite{soi3} \cite{soi2} is an integration-type SOI pixel detector. 
This detector has been utilized as an X-ray imaging device or charged particle detector in several applications, such as X-ray micro-computed tomography at a synchrotron radiation facility \cite{ryunishi1} \cite{ryunishi2} \cite{ryunishiD}, residual stress measurements \cite{cosa}, and beam tests on new devices \cite{sofist}  \cite{fpix}. 
The pixel size is 17 $\times$ 17 $\mathrm{\mu m^2}$, the number of pixels is 425,984 (column 832 $\times$ row 512), 
and the sensitive area is 14.1 $\times$ 8.7 $\mathrm{mm^2}$. 
The detector consists of 13 blocks (column 64 $\times$ row 512 pixels per block) 
and each block has independent channels of analog output for parallel readout.
A photograph of INTPIX4 is presented in Fig. \ref{fig:intpix4}. 

\subsection{Implementation of INTPIX4 for residual stress measurement system}

\begin{figure}[htb]
\centering
\includegraphics[width=\linewidth]{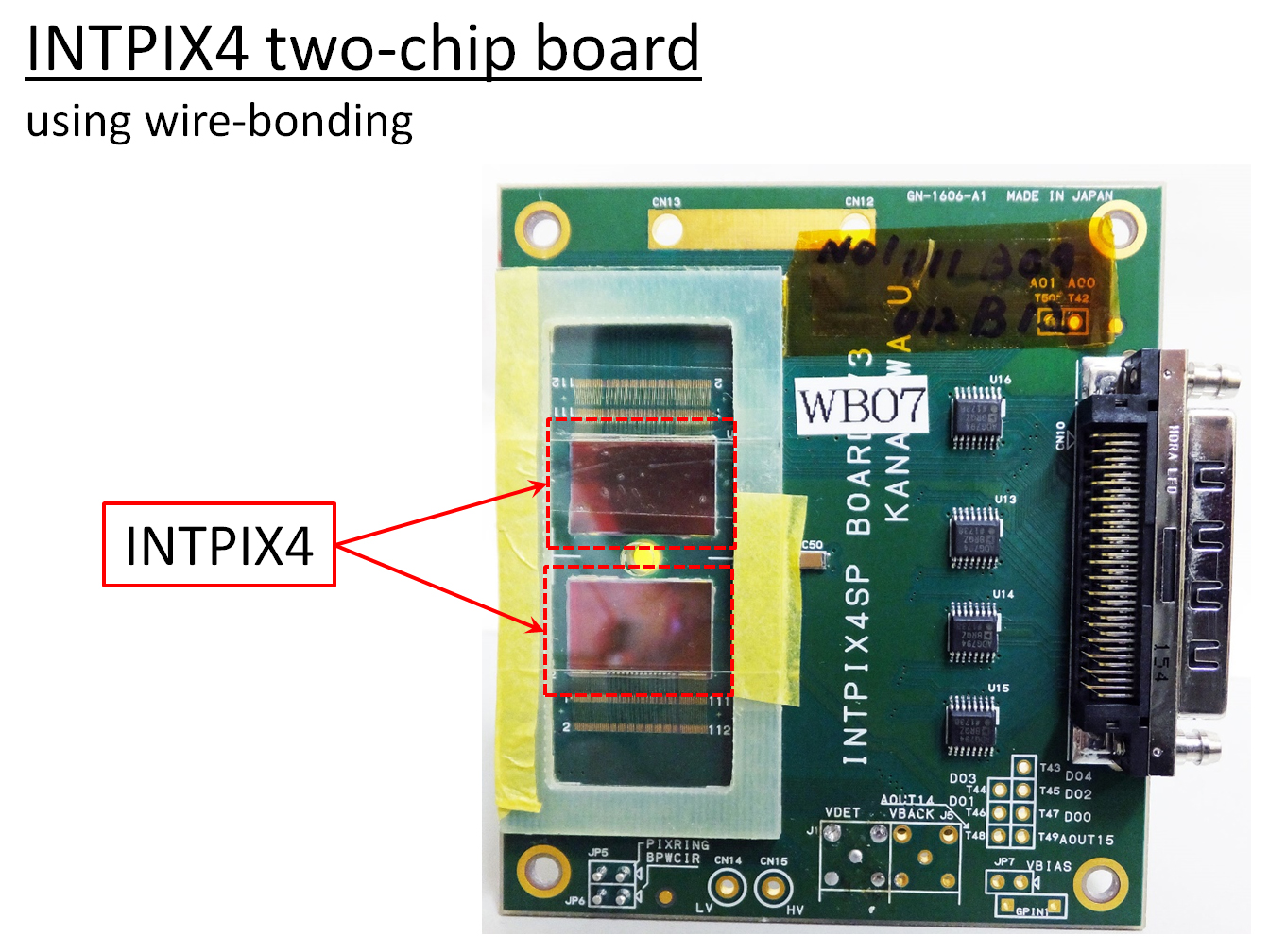}
\caption{Photograph of INTPIX4 two-chip board using wire bonding.}
\label{fig:intpix4wb}
\end{figure}

Two INTPIX4 chips are used in the residual stress measurement system. 
These chips are mounted on the sensor sub-board, which has a hole at the center, between the two chips, for the X-ray beam to pass through (Fig. \ref{fig:intpix4wb}).
The outside dimension of the sensor sub-board is 80 $\times$ 71 $\mathrm{mm^2}$, the edge-to-edge distance between the chips is typically 3 mm  (die) and 3.737 mm (sensitive area), and the diameter of the hole for the X-ray beam is 2.5 mm.
The sensor thickness of the implemented INTPIX4s is 300 $\mathrm{\mu m}$, the float zone wafer is N-type, and the typical reverse bias voltage of the sensor is 70 V. 
An Soi EvAluation BoArd with SiTCP \cite{seabas} version 2 (SEABAS2) and its DAQ system \cite{ryunishiD} \cite{ryunishi3} are used for readout and to operate INTPIX4. 
This system uses the SEABAS2 board as a DAQ platform board. 
The board has a 16-channel analog-to-digital converter, 
a 4-channel digital-to-analog converter, a field-programmable gate array, and a Gigabit Ethernet I/F. 

\subsection{Decision procedure for the shape of Debye ring}

\begin{figure}[htb]
\centering
\includegraphics[width=\linewidth]{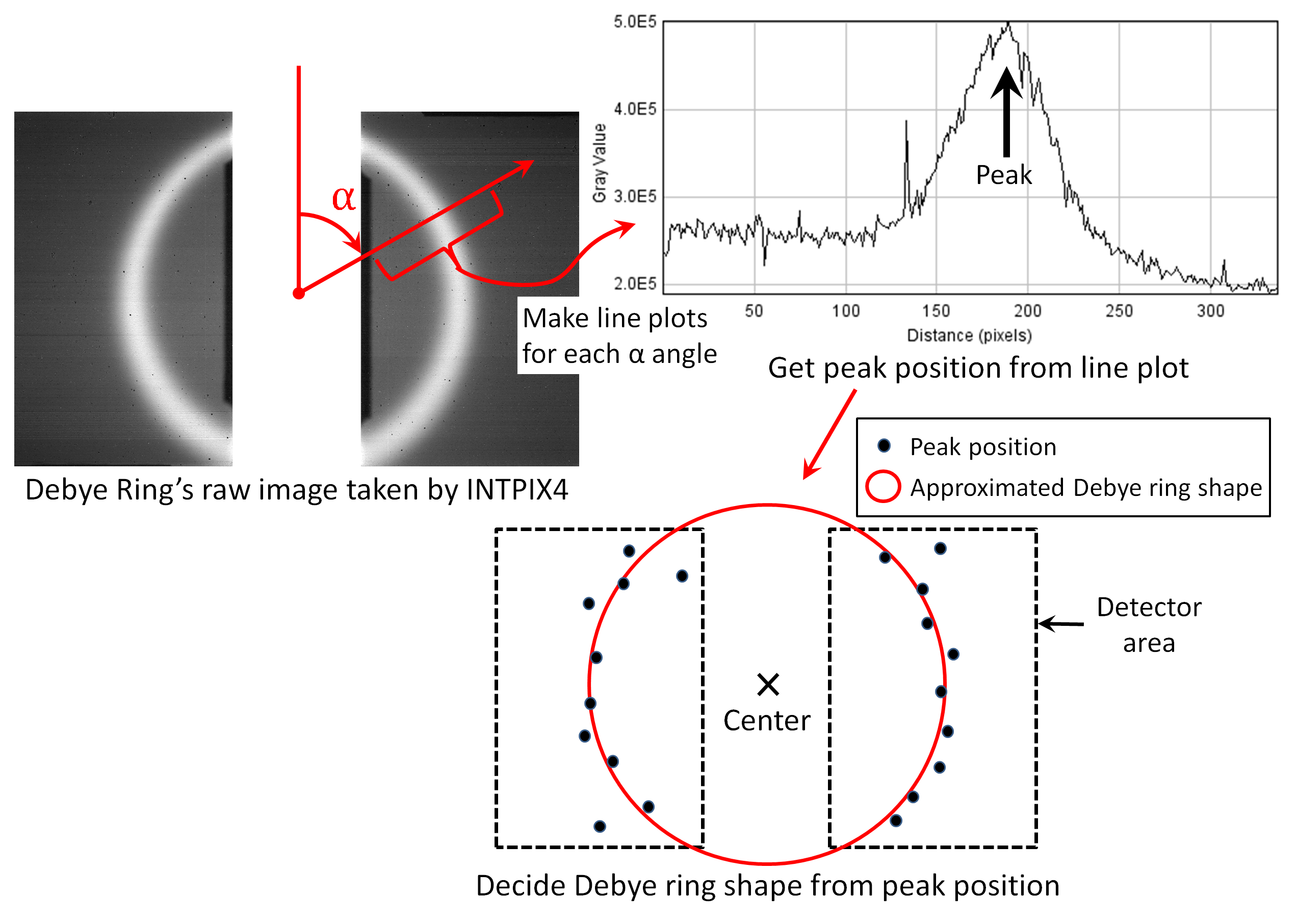}
\caption{Decision procedure for shape of Debye ring.}
\label{fig:cosa_formdet}
\end{figure}

The Debye ring is captured as a two-dimensional image by INTPIX4. 
First, the intensity line profiles from the ring center position (predetermined from a no-stress Debye ring image taken using the same setup) for each $\alpha$-angle are created for this image. 
Next, the peak position is calculated from these line profiles, and a 2D map of this peak position is created. 
Finally, the shape of the Debye ring is determined from the elliptic approximation of this 2D map. 
This procedure is presented in Fig. \ref{fig:cosa_formdet}. 

\subsection{$cos \alpha$ method}

\begin{figure}[htb]
\centering
\includegraphics[width=\linewidth]{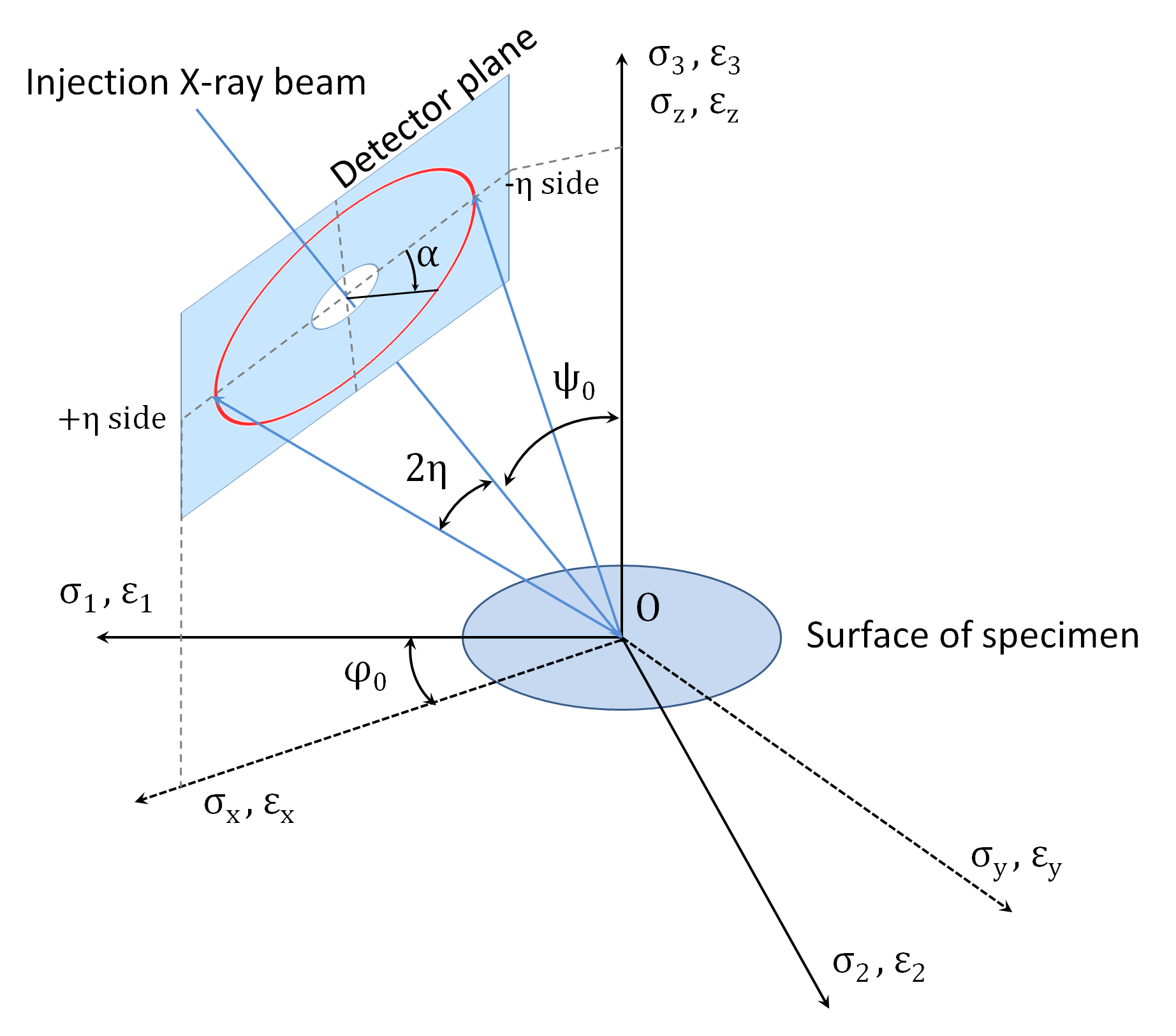}
\caption{Coordinate system of the $cos \alpha$ method. $\sigma_{1-3}$ are stress components and $\varepsilon_{1-3}$ are strain components in the sample coordinate system. $\sigma_{1-3}$ are stress components and $\varepsilon_{1-3}$ are strain components in the sample coordinate system. $\sigma_{x-z}$ are stress components and $\varepsilon_{x-z}$ are strain components in the experimental coordinate system. }
\label{fig:cosa_coodinate}
\end{figure}

\begin{figure}[htb]
\centering
\includegraphics[width=\linewidth]{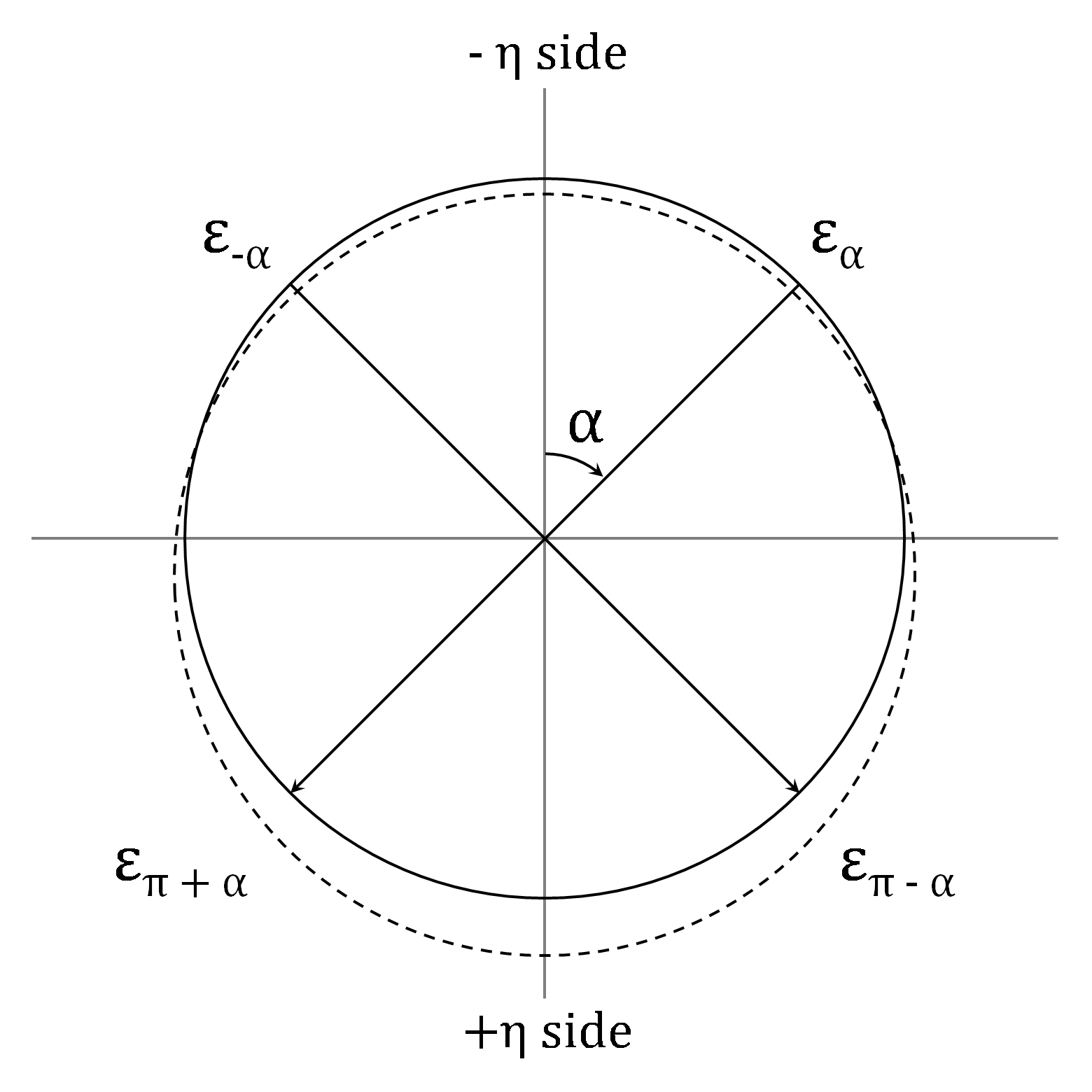}
\caption{Detector plane coordinate system of the $cos \alpha$ method.}
\label{fig:cosa_detplane}
\end{figure}

When residual stress occurs in metal, it appears as strains in the metal structure. 
These strains can be observed as the change in the lattice spacing $d$ of the metal from $d_{0}$ to $d_{0} + \Delta d$. 
When an X-ray beam is injected into the metal, the change in the lattice spacing $\Delta d$ is observed as the change in the Bragg angle $\theta$ from $\theta_{0}$ to $\theta_{0} + \Delta \theta$. 
The equation ${\Delta d}/{d_{0}} = -{\Delta \theta}/{tan \theta_{0}}$ is obtained by total differentiation of the Bragg formula 
$\lambda = 2d sin \theta$ (where $\lambda$ is the X-ray wavelength, which does not change in this situation). 
In the $cos \alpha$ method, the correspondence between the change in the lattice spacing $\Delta d$ due to strain 
and the Debye ring deformation, observable in the Debye ring image obtained by a 2D detector, is examined. 
The coordinate system used in the $cos \alpha$ method is illustrated in Fig. \ref{fig:cosa_coodinate}. 
The X-ray injection beam angle is defined as $\varphi_0$ and $\psi_0$. 
The angle $\alpha$ is determined in the clockwise direction from the beam injection side. 
The Debye ring deformation at angle $\alpha$ is $\varepsilon_\alpha$, expressed as 
\begin{equation}
\varepsilon_\alpha = -\frac{1}{tan\theta_0} \cdot (\theta - \theta_{0}),
\label{eq:cosa1}
\end{equation}
where $\theta_{0}$ is the Bragg angle prior to deformation (without strain) and $\theta =(\theta_{0} + \Delta \theta)$ is the Bragg angle following deformation. 
Typically, the X-ray penetration depth in a target metal material is several dozen micrometers from the metal surface. 
The measured value can be assumed to be the value under planar stress in such a situation. 
We introduce the variable $a_1$, defined as follows:
\begin{equation}
a_1 \equiv \frac{1}{2}\{(\varepsilon_\alpha - \varepsilon_{\pi + \alpha}) + (\varepsilon_{-\alpha} - \varepsilon_{\pi - \alpha})\},
\label{eq:cosa2}
\end{equation}
where 
$\varepsilon_\alpha$, $ \varepsilon_{\pi + \alpha}$, $\varepsilon_{-\alpha}$, and $\varepsilon_{\pi - \alpha}$ are four deformation points (Fig. \ref{fig:cosa_detplane}). Moreover, 
$a_1$ can be rewritten by converting the coordinates of the Debye ring to those of the metal surface: 
\begin{equation}
a_1 = -\frac{1+\nu}{E} \cdot sin 2\eta \cdot sin2\psi_0 \cdot cos\alpha \cdot \sigma_x.
\label{eq:cosa3}
\end{equation}
Here, $E$ is the X-ray Young's modulus, $\nu$ is the X-ray Poisson's ratio, 
and $\sigma_x$ is the $x$-component of the normal stress at the X-ray injection point. 
The relationship between $\eta$ and $\theta$ is obtained as $2\eta = \pi - 2\theta$. 
Partially differentiating both sides of formula (\ref{eq:cosa3}) with respect to $cos \alpha$ results in this expression for $\sigma_x$: 
\begin{equation}
\sigma_x = -\frac{E}{1+\nu} \cdot \frac{1}{sin 2\eta \cdot sin2\psi_0} \cdot (\frac{\partial a_1}{\partial cos\alpha}).
\label{eq:cosa4}
\end{equation}
Formula (\ref{eq:cosa4}) indicates that $\sigma_x$ can be calculated from the derivative of $a_1$ as a function of $cos \alpha$. 
Thus, $\sigma_x$ is determined according to the slope of the $cos \alpha$ -- $a_1$ plot. 
Furthermore, the y-components of the normal stress $\sigma_y$ and shearing stress $\tau_{xy}$ can be calculated in a similar way \cite{cosa3} \cite{cosa4} \cite{cosa5}.

\section{Experiment}

We attempted to measure the residual stress value of the specimen using the first prototype setup at the KEK Photon Factory BL-14A beam line. 
This experiment was designed to evaluate the fine accuracy and fine sampling pitch distribution for the next experiment. 

\subsection{Setup}

The setup consisted of a vacuum beam path (approximately 1.0--1.5 $\times$ $10^{-1}$ mBar, 42 cm), a wire-bonding-type two-chip SOI pixel detector board, stages, and a sample holder for the specimen. 
A vacuum beam path was installed to lead the X-ray beam without intensity loss due to air. 
The residual stress standard specimen for this experiment was produced by SINTOKOGIO, LTD. (Sinto).
The details of the beam configuration are presented in Table \ref{tab:exp_setup_detail}. 
The INTPIX4 detector settings are displayed in Table \ref{tab:exp_setup_detail_soi} and the specimen specifications are indicated in Table \ref{tab:exp_setup_detail_spe}. 
A photograph and schematic of the overall setup are presented in Fig. \ref{fig:exp_result}. 

\begin{table}[htb]
\caption{Beam configuration of experiment at Photon Factory.}
\begin{tabularx}{\linewidth}{XX}
\hline
Period & December 2019 \\
Location & Photon Factory BL-14A \\
PF ring (storage ring) mode & Top-up (HYBRID 30 mA + 420 mA) \\
X-ray energy & 5.415 keV monochromatic \\
Injection beam diameter & 2.5 mm \\
Beam intensity (estimation) & 4 $\times$ $10^{8}$ photons/s in entire injection beam \\
\hline
\end{tabularx}
\label{tab:exp_setup_detail}
\end{table}

\begin{table}[htb]
\caption{INTPIX4 detector operation parameters.}
\begin{tabularx}{\linewidth}{XX}
\hline
Detector & INTPIX4 (300 $\mathrm{\mu m}$ thickness, FZn-type wafer) \\
Sensor bias voltage & +200 V \\
Exposure time & 200 s (100 ms/frame $\times$ 2000 frames) \\
Scan time & 320 ns/pix \\
Sensor node reset & 4 $\mathrm{\mu s}$ with 300 mV ref. voltage \\
CDS reset & 5 $\mathrm{\mu s}$ with 350 mV ref. voltage \\
Readout board & SEABAS 2 \\
\hline
\end{tabularx}
\label{tab:exp_setup_detail_soi}
\end{table}

\begin{table}[htb]
\caption{Specimen specifications}
\begin{tabularx}{\linewidth}{XX}
\hline
Manufacturer & SINTOKOGIO, LTD. (Nagoya, Aichi, Japan) \\
Material & Steel \\
Diffraction plane & (h, k, l) = (2, 1, 1) \\
Nominal value & -384.0 MPa (average of 5 tests of center point) \\
Beam injection point calibrated by & Sinto Fe powder specimen as 0 MPa sample \\
\hline
\end{tabularx}
\label{tab:exp_setup_detail_spe}
\end{table}

\begin{figure}[htb]
\centering
\includegraphics[width=\linewidth]{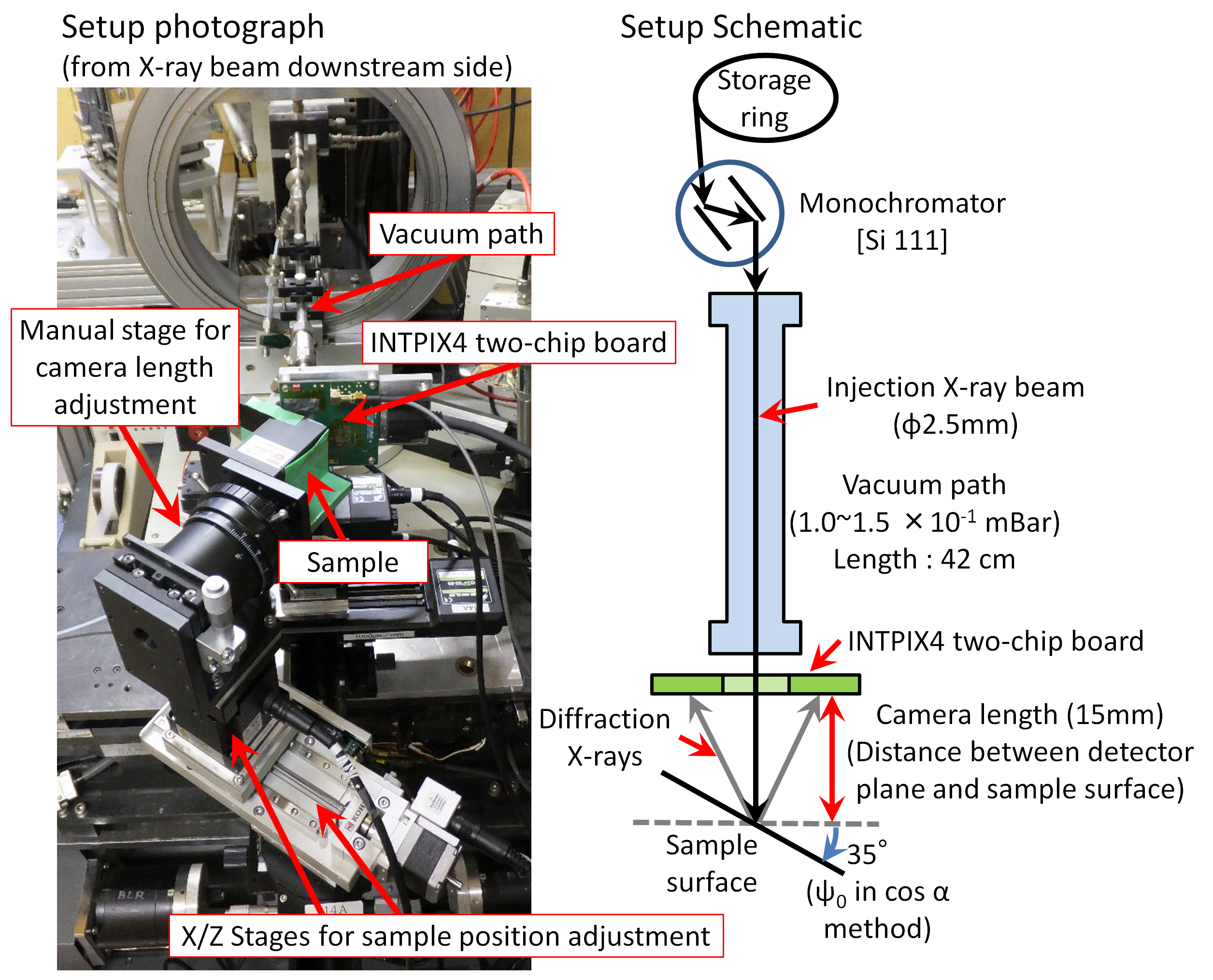}
\caption{Experimental setup photograph (left) and schematic (right).}
\label{fig:exp_setup}
\end{figure}

\subsection{Procedure of 2D mapping}

\begin{figure}[htb]
\centering
\includegraphics[width=\linewidth]{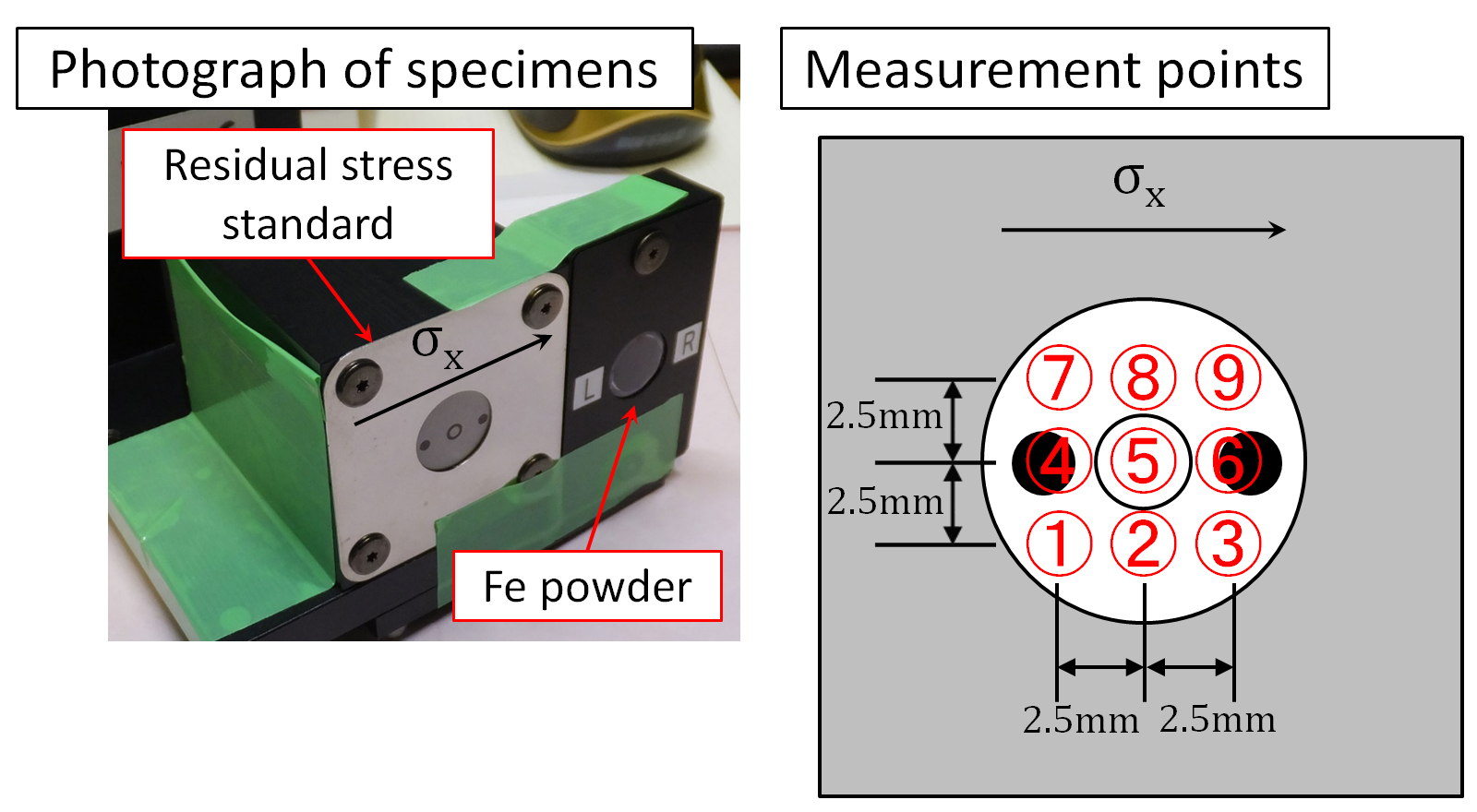}
\caption{2D mapping specimen setup photograph (left) and measured point (right).}
\label{fig:exp_2dscan}
\end{figure}

In this experiment, we aimed to measure the residual stress distribution of the standard specimen. We set nine (3 $\times$ 3) measurement position points, including the center of the surface. Each point had a 2.5 mm distance from its neighboring points in the column or row direction. 
The measurements proceeded as follows: 
\begin{enumerate}
\setlength{\itemsep}{1pt}
\item Measure Debye ring of Fe powder specimen (Sinto).
\item Measure Debye ring of residual stress standard specimen (Sinto)
(2.5 mm step, 3 $\times$ 3 points including the center point).
\item Calculate injection beam center from the Debye ring of Fe powder specimen.
\item Calculate distribution of residual stress amount on residual stress standard specimen from the Debye ring of each point.
\item Repeat steps 1 to 4.
\end{enumerate}

\subsection{Results}

\begin{figure}[htb]
\centering
\includegraphics[width=\linewidth]{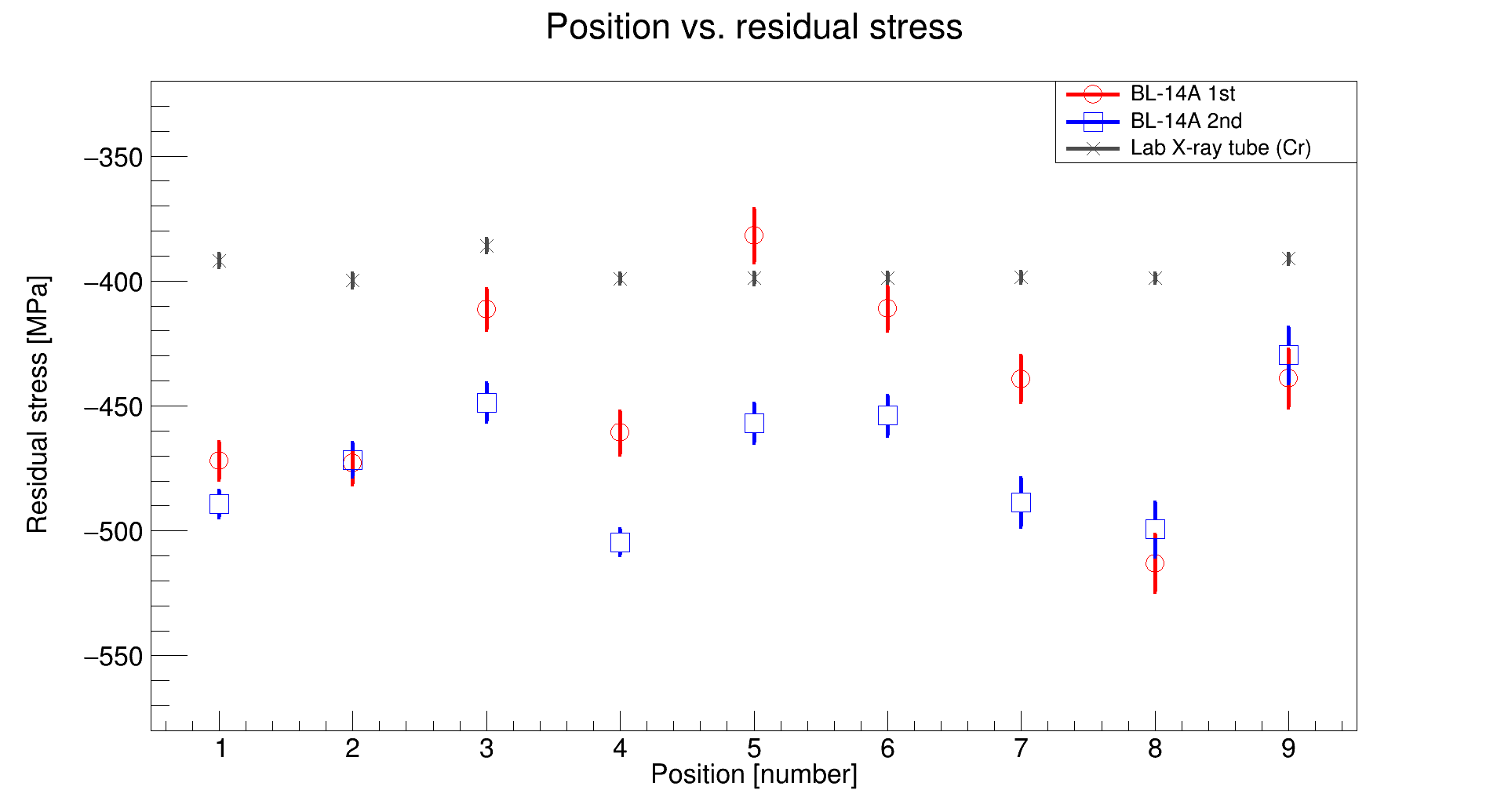}
\caption{Results of nine-point 2D mapping with residual stress standard specimen (nominal value: -384.0 MPa). 
The red circles (BL-14A 1st) indicate the first synchrotron measurement results and the blue squares (BL-14A 2nd) represent the second results (repeat measurements). 
The black crosses (Lab X-ray tube (Cr)) are the values measured by the laboratory X-ray tube (Cr K$\alpha$), included as a reference.}
\label{fig:exp_result}
\end{figure}

\begin{table}[htb]
\caption{Results of nine-point 2D mapping with residual stress standard specimen (nominal value: -384.0 MPa).}
\begin{tabularx}{\linewidth}{cXXX}
 & \multicolumn{3}{c}{Stress value (MPa)} \\
\multirow{2}{*}{Position} & \multicolumn{2}{l}{Synchrotron measurement} & \multirow{2}{*}{Laboratory measurement} \\
 & 1st & 2nd & \\
\hline
1 & -472.1 ($\pm$7.6) & -489.3 ($\pm$5.4) & -391.8 ($\pm$2.6) \\
2 & -473.1 ($\pm$8.2) & -471.5 ($\pm$6.8) & -399.7 ($\pm$2.7) \\
3 & -411.3 ($\pm$8.2) & -448.6 ($\pm$7.8) & -385.9 ($\pm$2.7) \\
4 & -460.8 ($\pm$8.6) & -504.6 ($\pm$5.2) & -399.1 ($\pm$2.0) \\
5 & -381.9 ($\pm$10.6) & -457.0 ($\pm$7.9) & -398.9 ($\pm$2.3) \\
6 & -411.0 ($\pm$8.8) & -453.9 ($\pm$8.0) & -398.7 ($\pm$2.1) \\
7 & -439.2 ($\pm$9.3) & -488.7 ($\pm$9.7) & -398.4 ($\pm$2.2) \\
8 & -513.0 ($\pm$11.5) & -499.4 ($\pm$10.8) & -398.7 ($\pm$1.9) \\
9 & -439.0 ($\pm$11.5) & -429.7 ($\pm$11.1) & -391.1 ($\pm$1.9) \\
\hline
\end{tabularx}
\label{tab:exp_result_list}
\end{table}

Figure \ref{fig:exp_result} presents a plot of the 2D mapping results and Table \ref{tab:exp_result_list} displays the values thereof, where the position numbers correspond to those in Fig. \ref{fig:exp_2dscan}.
The error values depicted in the results are estimations obtained from the linear approximation of the $cos \alpha$--$a_1$ plot. 
Thus, the error shown for one measurement includes the statistical uncertainty resulting from the dispersion of the peak position used in determining the Debye ring shape, but does not include the systematic uncertainty, such as that resulting from the instability of the setup. 

\subsection{Discussion}

Firstly, it can be observed that the synchrotron X-ray beam data exhibited errors that were larger and a fluctuation range that was wider than those of the laboratory data. 
The statistical uncertainty part of this error can probably be explained by the limited statistics caused by a lack of beam intensity and/or the detection efficiency of the detector. 
The real detection efficiency in this synchrotron measurement setup was 10 times lower than the estimation, based on the beam intensity as measured by the sodium iodide (NaI) scintillation detector and on the detection efficiency of the SOI detector in the specification. 
A possible explanation for this is that the X-ray signal was buried in the fluctuation of the thermal noise and dark current noise caused by operating INTPIX4 at room temperature. 
The analog signal output level of the INTPIX4 detector was approximately 15 mV per photon for 5.415 keV X-ray, and its noise level at room temperature typically fluctuated from 5 to 20 mV. 
The systematic uncertainty part of this error can probably be explained by the instability of the setup, caused by the rigidity of the stages or the sample holder. 
If the rigidity of the setup is insufficient, the specimen surface will tilt. 
This may affect the camera length (the distance between the detector and specimen surface) and $\psi_0$ angle. 
These parameters were used for the calculation of $\Delta \theta$ from the Debye ring deformation. 
If an error in these parameters results in a shift of the Debye ring deformation of several tens of a $\mathrm{\mu m}$, this will be equivalent to several tens of MPa of stress. 

Moreover, it can be observed that, in general, the synchrotron data exhibited 10 to 100 MPa stronger compressive stress values than the laboratory data. 
This can be explained by the different X-ray energy spectrum (the laboratory X-ray tube had no monochromator; therefore, its output was a white beam with a Cr K$\alpha$ peak). 
Higher X-ray energy will result in deeper penetration in the specimen. 
Therefore, the measured residual stress value will be distributed from the shallow to deeper regions of the metal. 
If the specimen has stress gradients in the depth direction \cite{doelle, torres}, the synchrotron data and laboratory data will exhibit different values. 

Another possible explanation is the difference in parallelism of the X-ray beams. 
The laboratory X-ray beam is a collimated cone beam. 
Therefore, its injection area on the specimen surface is wider than the injection area of the synchrotron beam. 
Furthermore, the laboratory X-ray beam has a wider dispersion of the incident angle. 

There was also a large difference between the first and second synchrotron measurements. 
In the laboratory environment, high reproducibility of stress values was confirmed by repeated measurements for the same position with the INTPIX4 detector setup \cite{mitsui}. 
A possible explanation for the differences in the synchrotron measurements is that they were caused by poor positional reproducibility resulting from the instability of the setup. 

\section{Conclusions}

X-ray residual stress measurement (the $cos \alpha$ method) is a non-destructive residual stress evaluation technique using the Debye ring deformation. 
In this study, we attempted to measure the residual stress using a synchrotron monochromatic X-ray beam (5.415 keV) at Photon Factory BL-14A, and we obtained initial results with the first prototype setup. 
Furthermore, we compared our synchrotron measurements with laboratory measurements. 
The synchrotron X-ray beam data exhibited larger errors, probably caused by limited statistics and instability of the setup. 
Although the cause of the limited statistics was not identified, one possibility was the detector's detection efficiency. 
We will continue to investigate the true detection efficiency of the detector and improve the operating conditions of the detector. 
The instability of the setup was probably due to the rigidity of the stages or sample holder. 
Therefore, the stages or sample holder will be fixed in future experiments. 
However, even after considering the larger error, the synchrotron data exhibited 10 to 100 MPa stronger compressive stress values than the laboratory data. 
This can be explained by the difference in the X-ray source characteristics and the difference in the residual stress values according to the depth from the metal surface. 
To understand the actual cause, we will conduct several additional experiments in the future. 
For example, we will conduct a repeating measurement utilizing a small X-ray energy step (such as 0.1 to 0.5 keV/step from 5.4 keV) or one where we scrape  the surface of the specimen by means of electropolishing. 

\section{Acknowledgements}

This study was conducted under the approval of the Photon Factory Program Advisory Committee (Proposal No. 2019G606). 
The study was performed as part of the SOIPIX group's \cite{soipix} research activity. 
The study was supported by the ISIJ Research Promotion Grant from the Iron and Steel Institute of Japan. 
Development of INTPIX4 was supported by Systems Design Lab (d.lab), the University of Tokyo in collaboration with Cadence Design Systems, Inc., Synopsys, Inc., and Menter Graphics, Inc.
We would like to thank Editage (www.editage.com) for English language editing.

\end{document}